\documentclass[twocolumn,letterpaper]{revtex4}
\usepackage{graphicx}

\begin{document}
\title{
Room-temperature biphoton source with a spectral brightness near the ultimate limit
}

\author{
Jia-Mou Chen,$^1$ 
Chia-Yu Hsu,$^1$ 
Wei-Kai Huang,$^1$
Shih-Si Hsiao,$^1$ 
Fu-Chen Huang,$^1$ 
Yi-Hsin Chen,$^{2,5}$
Chih-Sung Chuu,$^{1,5}$ 
Ying-Cheng Chen,$^{3,5}$
Yong-Fan Chen,$^{4,5}$
and
Ite A. Yu$^{1,5,}$
}
\email{yu@phys.nthu.edu.tw}

\affiliation{
$^1$Department of Physics, National Tsing Hua University, Hsinchu 30013, Taiwan \\
$^2$Department of Physics, National Sun Yat-Sen University, Kaohsiung 80424, Taiwan \\
$^3$Institute of Atomic and Molecular Sciences, Academia Sinica, Taipei 10617, Taiwan\\
$^4$Department of Physics, National Cheng Kung University, Tainan 70101, Taiwan \\
$^5$Center for Quantum Technology, Hsinchu 30013, Taiwan
}

\begin{abstract}
The biphotons, generated from a hot atomic vapor via the process of spontaneous four-wave mixing (SFWM), have the following merits: stable and tunable frequencies as well as linewidth. Such merits are very useful in the applications of long-distance quantum communication. However, the hot-atom SFWM biphoton sources previously had far lower values of generation rate per linewidth, i.e., spectral brightness, as compared with the sources of biphotons generated by the spontaneous parametric down conversion (SPDC) process. Here, we report a hot-atom SFWM source of biphotons with a linewidth of 960 kHz and a generation rate of 3.7$\times$$10^5$ pairs/s. The high generation rate, together with the narrow linewidth, results in a spectral brightness of 3.8$\times$$10^5$ pairs/s/MHz, which is 17 times of the previous best result with atomic vapors and also better than all known results with all kinds of media. The all-copropagating scheme together with a large optical depth (OD) of the atomic vapor is the key improvement, enabling the achieved spectral brightness to be about one quarter of the ultimate limit. Furthermore, this biphoton source had a signal-to-background ratio (SBR) of 2.7, which violated the Cauchy-Schwartz inequality for classical light by about 3.6 folds. Although an increasing spectral brightness usually leads to a decreasing SBR, our systematic study indicates that both of the present spectral brightness and SBR can be enhanced by further increasing the OD. This work demonstrates a significant advancement and provides useful knowledge in the quantum technology using photons.
\end{abstract}

\maketitle

\newcommand{\FigOne}{
	\begin{figure}[t]
	\center{\includegraphics[width=87mm]{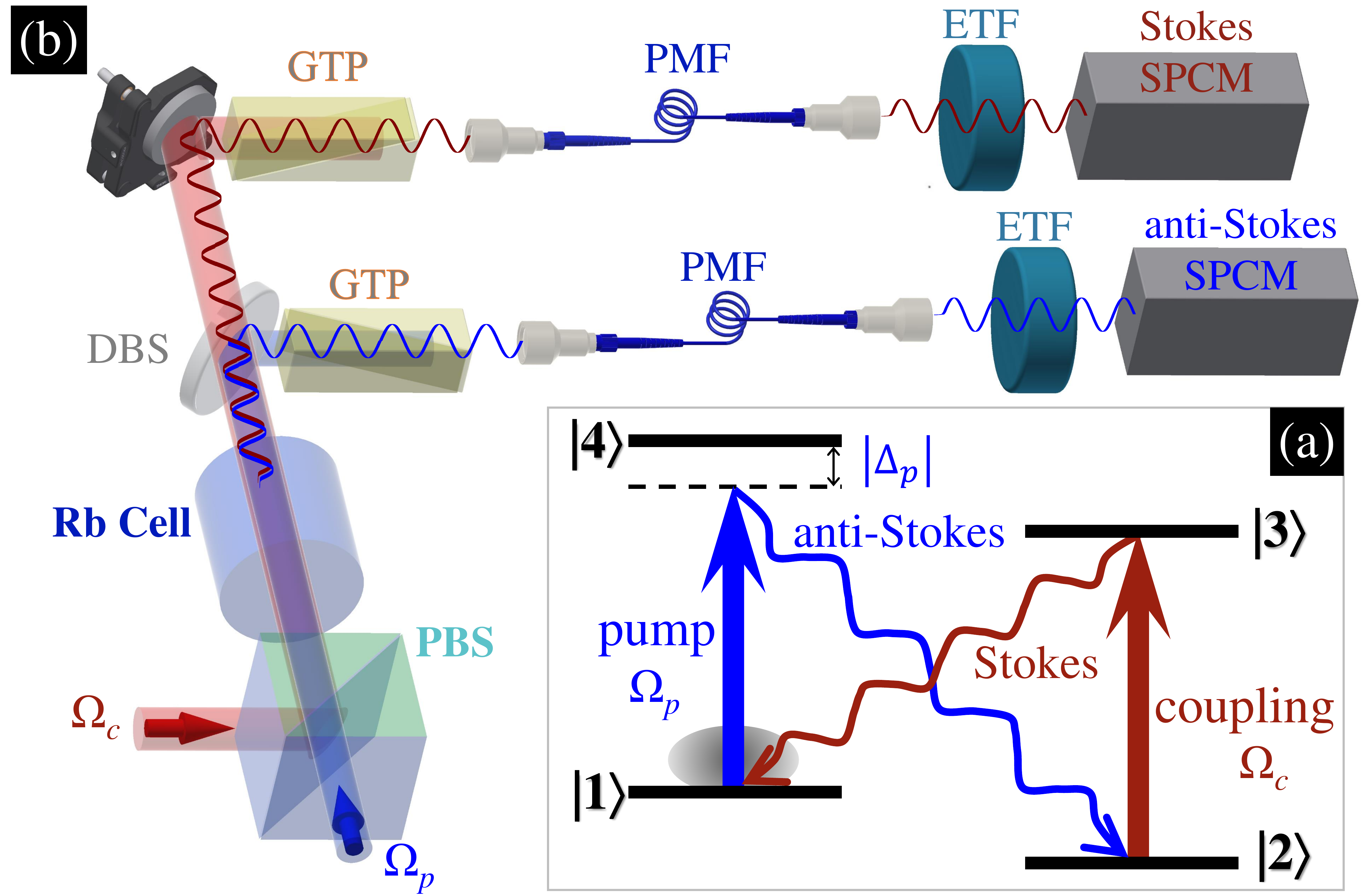}}
	\caption{(a) Relevant energy levels and transition scheme in the experiment. $|1\rangle$, $|2\rangle$, $|3\rangle$, and $|4\rangle$ represent the ground states of $|5S_{1/2},F=2\rangle$ and $|5S_{1/2},F=1\rangle$ and the excited states of $|5P_{1/2},F=2\rangle$ and $|5P_{3/2},F=1,2\rangle$ of $^{87}$Rb atoms, respectively. The wavelengths of the pump field and anti-Stokes photons are about 780 nm, and those of the coupling and Stokes photons are about 795 nm. Throughout this study, the pump detuning, $\Delta_p$, was set to $-2.0$ GHz with respect to the transition of $|1\rangle \rightarrow |5P_{3/2},F=2\rangle$, and the coupling field frequency was tuned to the resonance. 
	(b) Schematic experimental setup. PBS: polarizing beam splitter, DBS: dichroic  beam splitter, GTP: Glan-Thompson polarizer, PMF: polarization-maintained optical fiber, ETF: etalon filter, and SPCM: single-photon counting module.}
	
	\label{fig:one}
	\end{figure}
}
\newcommand{\FigTwo}{
	\begin{figure}[t]
	\center{\includegraphics[width=70mm]{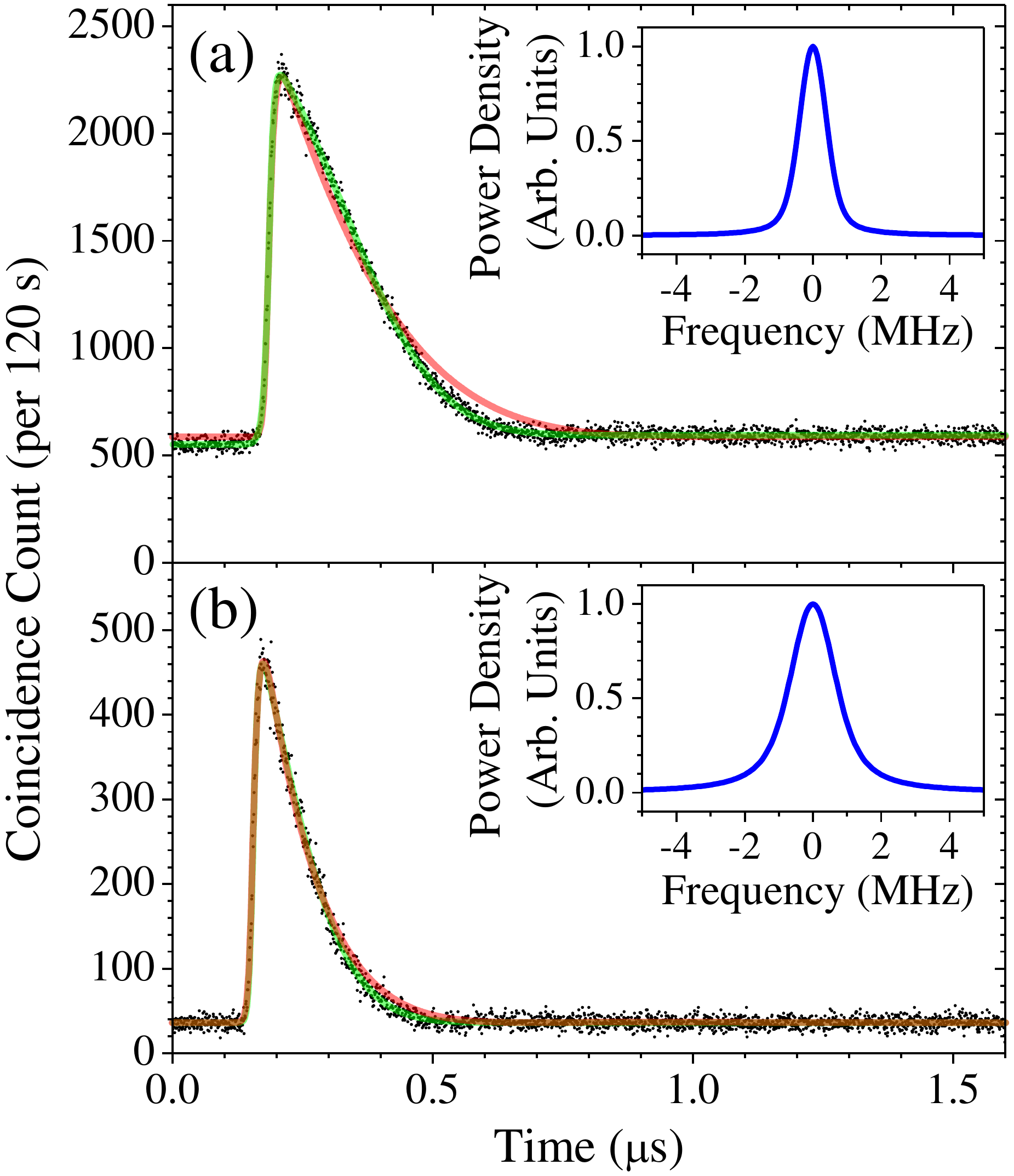}}
	\caption{Coincidence count of the anti-Stokes and Stokes photons as a function of the delay time between them is shown in each main plot. Black dots represent numbers of coincidence counts per time bin accumulated in 120 s, where the width of time bin is 0.8 ns. The pump and coupling powers are 16 and 4 mW. At the vapor cell temperatures of 65 and 38 $^{\circ}$C, the detection rates in (a) and (b) are 4.0$\times$10$^3$ and 560 pairs/s, respectively, corresponding to the generation rates of 3.7$\times$10$^5$ and 5.1$\times$10$^4$ pairs/s. Green and red lines are the best fits and theoretical predictions. In the theoretical calculation, $\alpha =$ 370 in (a) or 93 in (b), $\Omega_c$ = 5.4$\Gamma$, and $\gamma =$ 0.030$\Gamma$ in (a) or 0.020$\Gamma$ in (b). The full width at the half maximum (FWHM) of the best fit in (a) is 180 ns and that in (b) is 100 ns. In the insets, blue lines are the Fourier transforms of the best fits, and their FWHMs are 960 kHz and 1.6 MHz. 
	}
	\label{fig:two}
	\end{figure}
}
\newcommand{\FigThree}{
	\begin{figure}[t]
	\center{\includegraphics[width=85mm]{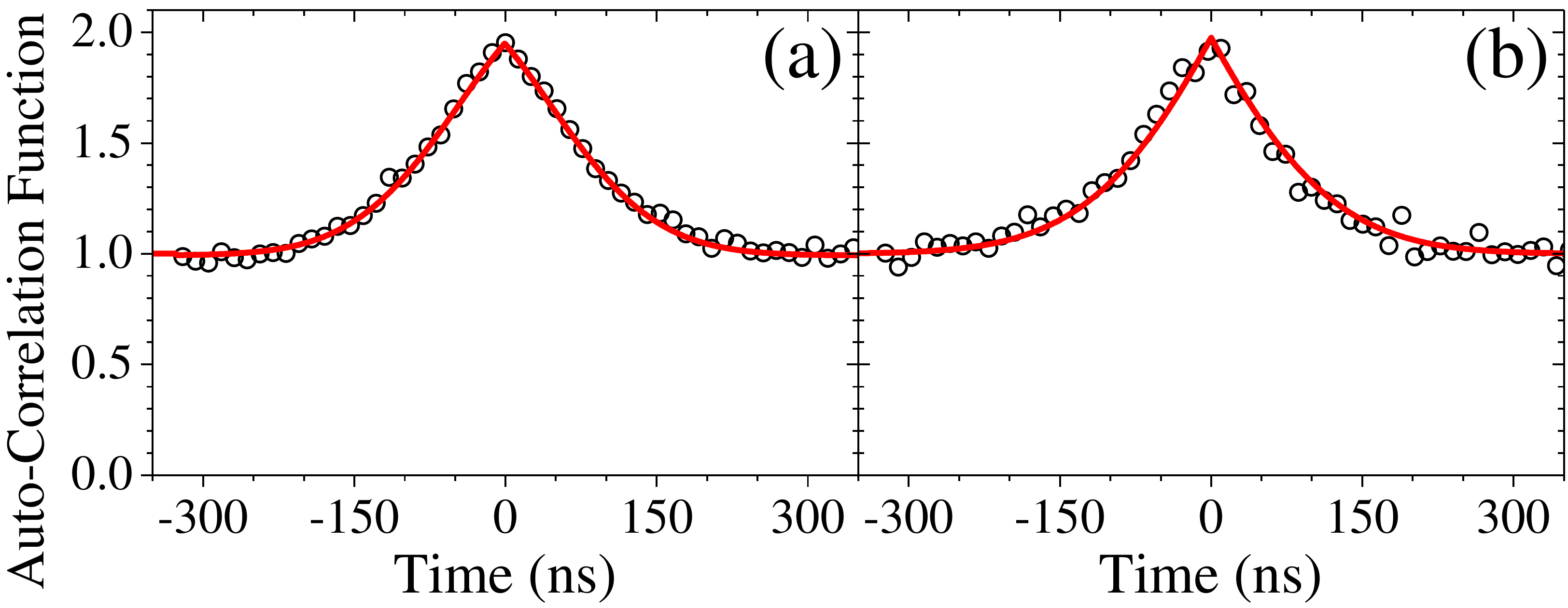}}
	\caption{(a) The auto-correlation function of anti-Stokes photons, $g^{(2)}_{as,as}$, and (b) that of Stokes photons, $g^{(2)}_{s,s}$, as functions of the delay time. Circles are the experimental data, measured at the condition of the highest generation rate whose representative data are shown in Fig.~\ref{fig:two}(a). The accumulation time was 120 s and the width of time bin was 12.8 ns. Red lines are the best fits, which determine (a) $g^{(2)}_{as,as}(0) =$ 1.95$\pm$0.03 and the FWHM of 150~ns, and (b) $g^{(2)}_{s,s}(0) =$ 1.97$\pm$0.07 and the FWHM of 140~ns.
	}
	\label{fig:newthree}
	\end{figure}
}
\newcommand{\FigFour}{
	\begin{figure}[t]
	\center{\includegraphics[width=85mm]{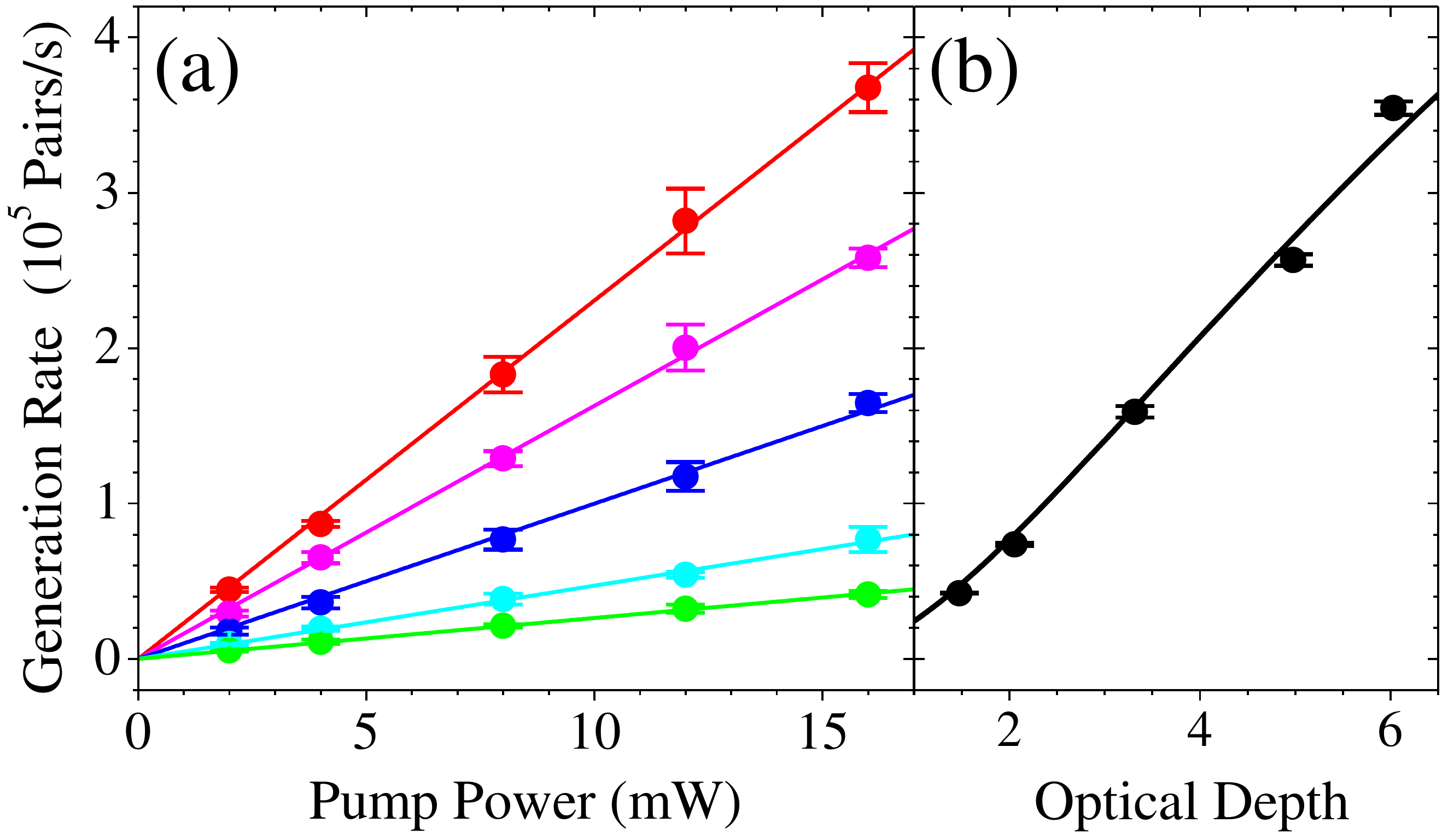}}
	\caption{(a) Generation rate as a function of the pump power. In all the measurements, we set the coupling powers to 4 mW. Green, cyan, blue, magenta, and red circles are the experimental data measured at the vapor cell temperatures of 38, 44, 53, 60, and 65 $^{\circ}$C. Lines are the best fits. (b) Black circles are the best fit in (a) at the pump power of 16 mW versus the measured optical depth, $\alpha'$, of each temperature. We only consider the fitting uncertainty in error bars. Black line is the theoretical prediction which is proportional to $\int G^{(2)}(\tau) d\tau$, where $G^{(2)}(\tau)$ is given by Eq.~(\ref{eq:biphoton}). In the calculation of the prediction, $\Omega_c$ = 5.4$\Gamma$, $\gamma =$ 0.025$\Gamma$, and $\alpha$ used in Eqs.~(\ref{eq:biphoton})-(\ref{eq:rho}) relates to $\alpha'$ according to Eq.~(\ref{eq:OD}).
	}
	\label{fig:three}
	\end{figure}
}
\newcommand{\FigFive}{
	\begin{figure}[t]
	\center{\includegraphics[width=85mm]{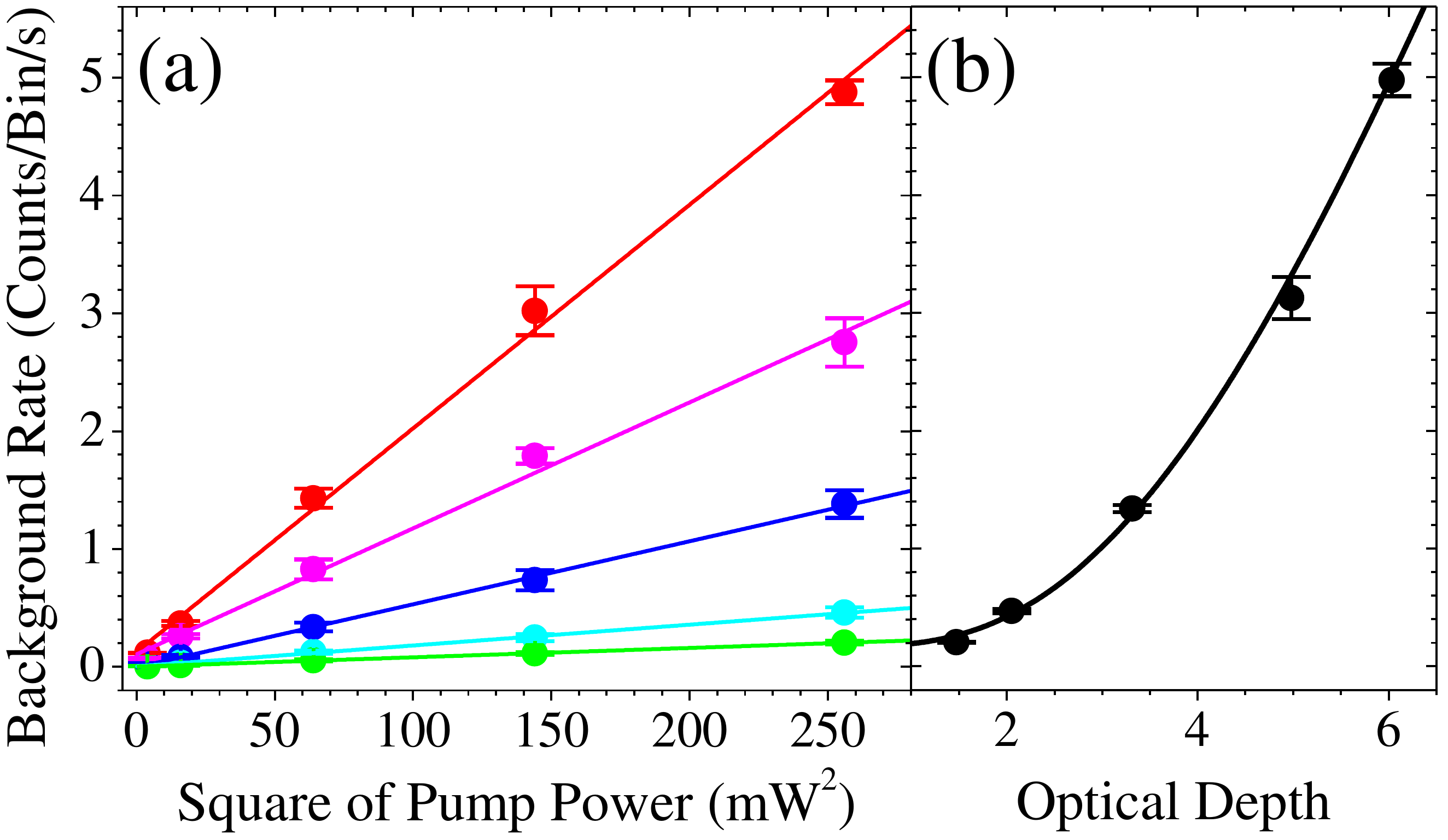}}
	\caption{(a) Rate of the background count per time bin as a function of the square of the pump power. The biphoton data and legends here are the same as those in Fig.~\ref{fig:three}. Lines are the best fits of straight lines with nonzero interceptions. (b)Black circles are the best fit in (a) at the pump power of 16 mW versus the measured optical depth, $\alpha'$, of each temperature. We only consider the fitting uncertainty in the error bars. Black line is the theoretical prediction which is proportional to $[\int G^{(2)}(\tau) d\tau]^2$, where $G^{(2)}(\tau)$ is given by Eq.~(\ref{eq:biphoton}). The calculation parameters of the prediction are the same as those in Fig.~\ref{fig:three}.
	}
	\label{fig:four}
	\end{figure}
}
\newcommand{\FigSix}{
	\begin{figure}[t]
	\center{\includegraphics[width=60mm]{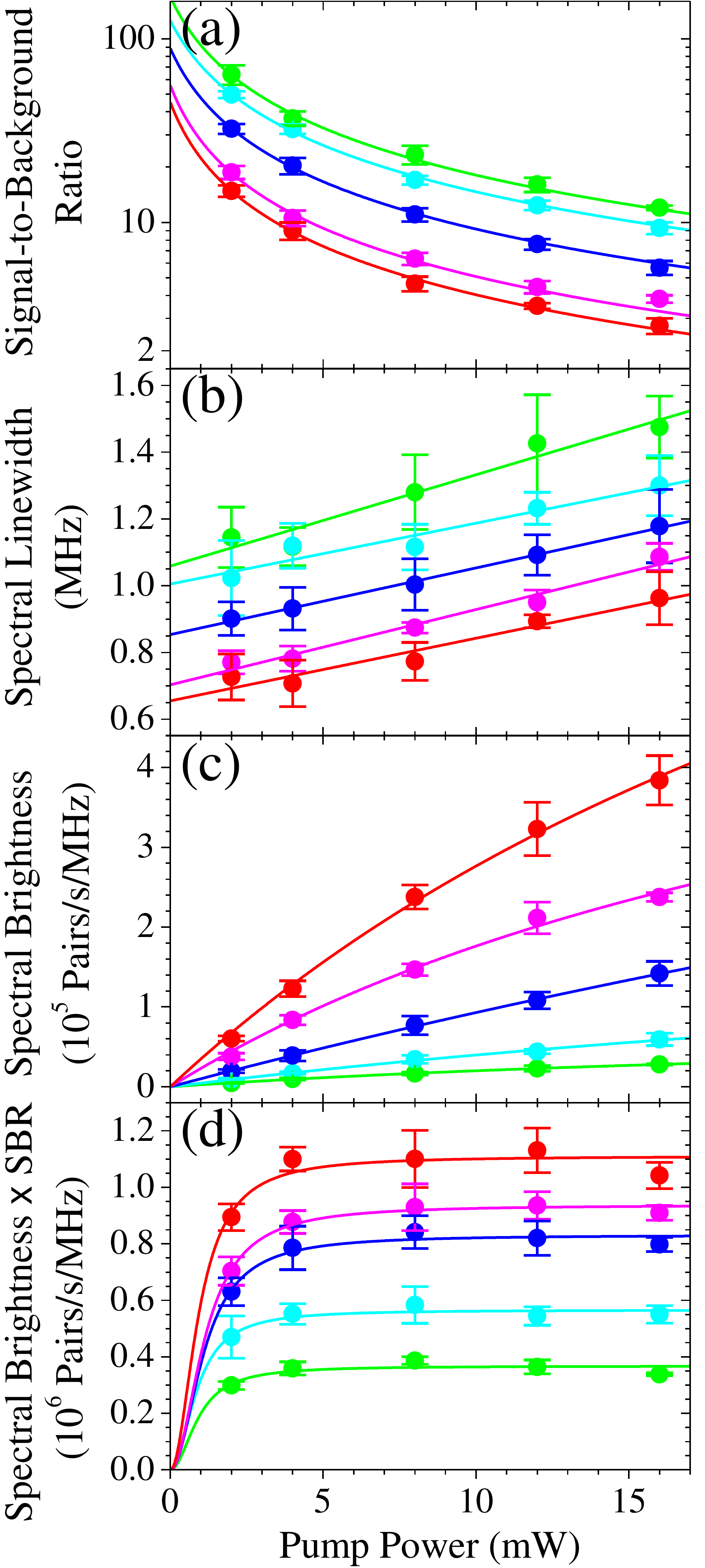}}
	\caption{Characteristics of the biphoton source as functions of the pump power: (a) signal-to-background ratio (SBR) in the logarithmic scale, (b) spectral linewidth, (c) spectral brightness, and (d) product of spectral brightness and SBR (denoted as $S$). In all the measurements, we set the coupling powers to 4 mW. Green, cyan, blue, magenta, and red circles are the experimental data taken at 38, 44, 53, 60, and 65~$^{\circ}$C. Lines are the best fits, and Eqs.~(\ref{eq:fit5a}), (\ref{eq:fit5b}), (\ref{eq:fit5c}), and (\ref{eq:fit5d}) are the fitting functions for the data in (a), (b), (c), and (d), respectively.
	}
	\label{fig:five}
	\end{figure}
}
\newcommand{\FigSeven}{
	\begin{figure}[t]
	\center{\includegraphics[width=75mm]{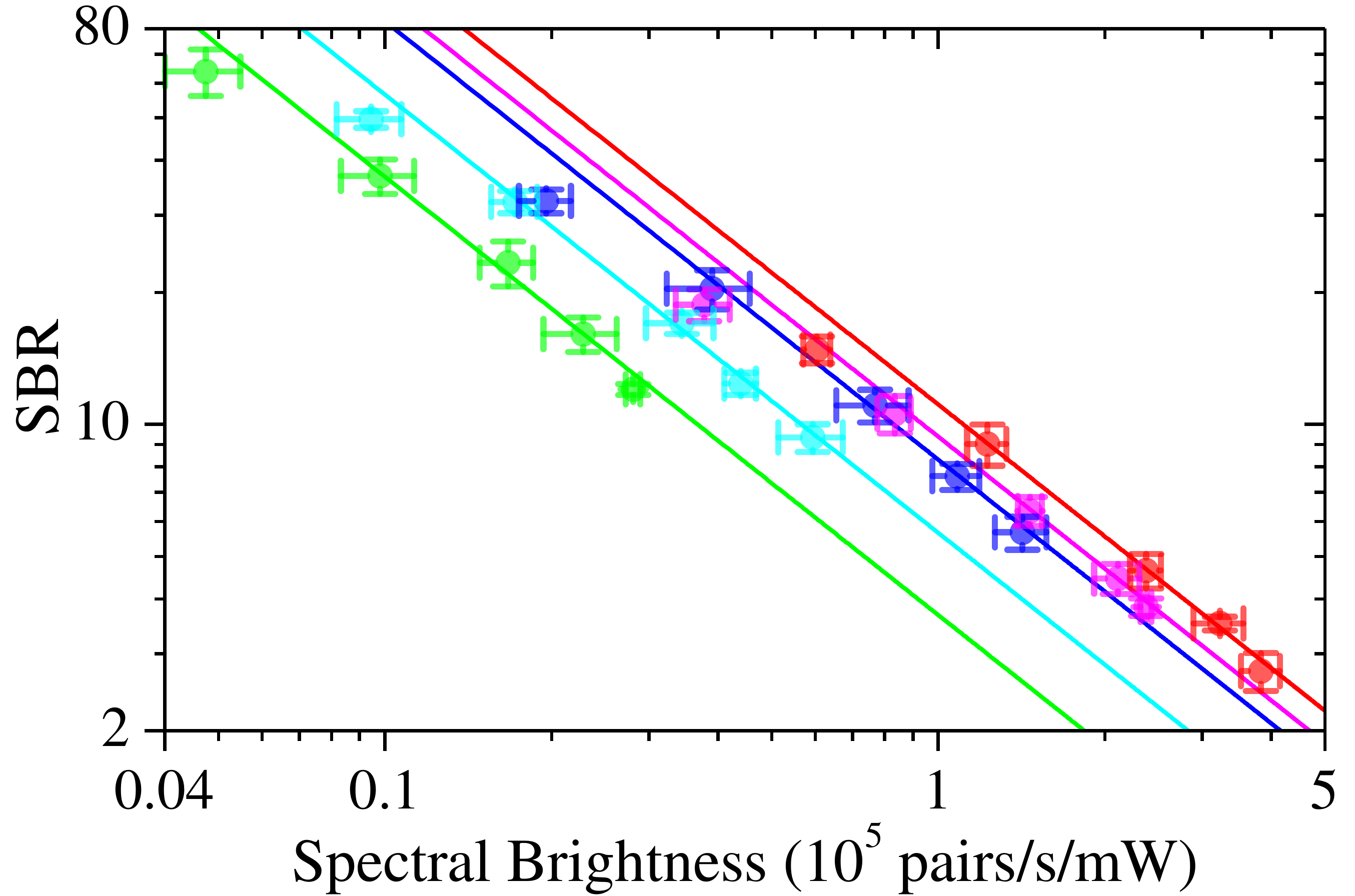}}
	\caption{The SBR is plotted against the spectral brightness.  Green, cyan, blue, magenta, and red circles are the experimental data taken at 38, 44, 53, 60, and 65~$^{\circ}$C. Each line represents the function of SRB $\times$ (spectral brightness) = $C$, where $C$ is the asymptotic limit, i.e., $A_4$ in Eq.~(\ref{eq:fit5d}), of the best fit in Fig.~\ref{fig:five}(d).
	}
	\label{fig:seven}
	\end{figure}
}
\section{Introduction}

Biphotons are pairs of time-correlated single photons, and the sum frequency and delay time of the two photons are entangled \cite{TEentanglement}. One photon of a pair can trigger or start an operation of a quantum process or device, and then the other photon in the same pair will be employed in the operation. The two photons are regarded as the heralding and heralded single photons, which are convenient and versatile in quantum information manipulation \cite{QI1, QD1, QD2, QD3, QD4}. Therefore, biphotons have been widely utilized in applications, such as quantum memory, quantum communication, quantum interference, and entanglement generation \cite{QM1, QM2, QM3, QM4, QC1, QC2, QC3, interference1, interference2, interference3, Entangle1, Entangle2}. 

Two major physical mechanisms are used to produce biphotons: the spontaneous parametric down conversion (SPDC) \cite{SPDCReview1, SPDCReview2, SPDC.Theory, brightSPDC5, brightSPDC6, narrowSPDC1, narrowSPDC2, narrowSPDC3, narrowSPDC4, brightSPDC1, brightSPDC2, brightSPDC3, brightSPDC4} and spontaneous four-wave mixing (SFWM) \cite{SFWMReview1, SFWMReview2, SFWM.Theory, narrowSFWM1, narrowSFWM2, narrowSFWM3, ColdAtoms1, ColdAtoms2, ColdAtoms3, ColdAtoms4, SFWM.dLambda.HotAtom1, SFWM.dLambda.HotAtom2, SFWM.dLambda.HotAtom3, SFWM.dLambda.HotAtom4, SFWM.dLambda.HotAtom5, SFWM.dLambda.HotAtom6, OurOPEX2021}. The SPDC is often employed for nonlinear crystals. An SPDC biphoton source can have a generation rate larger than 10$^6$ pairs/s. With the assistance of an optical cavity, the linewidth of SPDC biphotons can be narrowed down to a few MHz \cite{brightSPDC5, brightSPDC6, narrowSPDC1, narrowSPDC2, narrowSPDC3, narrowSPDC4}. The SFWM is often used for cold or hot atomic vapors. A cold-atom SFWM source can produce biphotons with a sub-MHz linewidth, because of the low decoherence rate in the cold-atom system \cite{narrowSFWM1, narrowSFWM2, narrowSFWM3}. However, the cold-atom SFWM source requires a timing sequence to switch the magneto-optical trap off and on, before and after the biphoton generation. Considering the duty cycle of the timing sequence, the SFWM biphoton source of cold atoms typically has a generation rate in the order of $10^{2}$ pairs/s \cite{ColdAtoms1, ColdAtoms2, ColdAtoms3, ColdAtoms4}.  

The SFWM biphoton source of hot atoms has the following merits. The frequency of its biphotons, being locked to lasers' frequencies due to the SFWM transition scheme, are more stable than the frequency of the SPDC biphotons, being locked to an optical cavity's temperature. The frequency of the hot-atom SFWM photons is continuously tunable in a range of 0.6 GHz or larger, based on the Doppler width of the atoms. This tuning range is comparable to that of the SPDC biphoton source, and it is significantly larger than that of the cold-atom SFWM biphoton source. The linewidth of the hot-atom SFWM biphotons can be varied for more than an order of magnitude and narrowed down to 290 kHz \cite{OurOPEX2021}, while the linewidth of SPDC biphotons is fixed by the optical cavity. Thus, the SFWM biphoton source of hot atoms is versatile in the application of long-distance quantum communication.

A biphoton source of a larger generation rate can produce information carriers faster to enable a higher bandwidth in the information transmission. A quantum operation utilizing biphotons of a narrower linewidth as information carriers can usually achieve a better efficiency. Thus, the generation rate per linewidth, i.e., the spectral brightness, is the combined measure of the generation rate and linewidth. The spectral brightness is an important figure of merit of a biphoton source, and it strongly influences the success rate of quantum communication. Among the single-mode SPDC biphoton sources, Luo {\it et al}.\ reported a 66-MHz biphoton source with the spectral brightness of 3$\times$ $10^5$ pairs/s/MHz \cite{brightSPDC2}, while Tsai {\it et al}.\ reported a 6-MHz biphoton source with the spectral brightness of 2.2$\times$ $10^5$ pairs/s/MHz \cite{brightSPDC3}. As for the hot-atom SFWM biphoton sources, Zhu {\it et al}.\ reported a 2.9-MHz biphoton source with the spectral brightness of 1.4$\times$ $10^4$ pairs/s/MHz \cite{SFWM.dLambda.HotAtom2}, and Hsu {\it et al}.\ reported a 630-kHz biphoton source with the spectral brightness of 2.3$\times$ $10^4$ pairs/s/MHz \cite{OurOPEX2021}. Prior to this work, the hot-atom SFWM biphoton source had a far lower spectral brightness, compared with the SPDC biphoton source. 

The biphoton sources made of integrated photonics devices with micro-resonators or waveguides have been developed recently \cite{IPD1, IPD2, IPD3, IPD4, IPD5, IPD6, IPD7, IPD8, IPD9, IPD10, IPD11, IPD12, IPDexample, IPD.Review1, IPD.Review2, IPD.Review3}. Either the SFWM or SPDC process can be employed in the generation scheme. Such a biphoton source has a large generation rate due to the high $Q$-factor of a resonator or the strong mode confinement of a waveguide, and also a high signal-to-background ratio (SBR) or coincidence-to-accidental ratio (CAR) \cite{IPD.Review1, IPD.Review2, IPD.Review3, background}. The linewidth of the biphotons emitted from the device can be as narrow as about 100~MHz. In Ref.~\onlinecite{IPDexample}, a very large generation rate of 1.3$\times$10$^7$ pairs/s at a high SBR or CAR of about 400 was achieved with a microring-resonator biphoton source, and the biphotons had a linewidth of 92 MHz. Considering the generation rate and the linewidth, the spectral brightness of this biphoton source was 1.4$\times$10$^5$ pairs/s/MHz.

We introduced the all-copropagating scheme to develop a hot-atom SFWM source. In the all-copropagating scheme as shown in Fig.~\ref{fig:one}, the pump and coupling fields and the anti-Stokes and Stokes photons all propagate in the same direction and completely overlap. The scheme is nearly free of phase mismatch and ensures a low decoherence rate. The phase match makes the generation process efficient, and the low decoherence rate results in a narrow biphoton linewidth. Thus, a high-rate source of narrow-linewidth biphotons was achieved in this work. At a vapor cell temperature of 65~$^\circ$C, the source of the 960-kHz biphotons in this work had a generation rate of 3.7$\times$ $10^5$ pairs/s, which is an order of magnitude better than the previous best result of the hot-atom SFWM source. The high generation rate, together with the narrow linewidth, gave a spectral brightness of 3.8$\times$10$^5$ pairs/s/MHz, which is 17 times of the previous best result with atomic vapors and also better than all known results with all kinds of media. 

\FigOne

The spectral brightness has an ultimate limit, which is explained as follows: The generation rate is equal to $1/\tau_d$, where $\tau_d$ is the average separation time between pairs. While $\tau_p$ is the biphoton's temporal width, the linewidth is equal to $1/(2\pi\tau_p)$. Hence, the spectral brightness or the generation rate divided by the linewidth is given by $2\pi\tau_p/\tau_d$. If the value of $2\pi\tau_p/\tau_d$ gets too large, biphotons often overlap their wave packets and they are no longer nonclassical. Consequently, the spectral brightness has an upper limit to ensure that biphotons are reasonably separated in time and preserve their nonclassicality. We estimate the limit by using the condition of $\tau_p/\tau_d =$ 0.25, i.e., once a pair is generated, the probability of another pair generated within a time interval of $\tau_p$ is 0.25. This limit is given by $(\pi/2)$$\times$10$^6$ pairs/s/MHz. The highest spectral brightness in this work is 3.8$\times$10$^5$ counts/s/MHz, which is about one quarter of the ultimate limit.

The counter-propagating scheme was commonly used in the previous hot-atom SFWM sources \cite{SFWM.dLambda.HotAtom1, SFWM.dLambda.HotAtom2, SFWM.dLambda.HotAtom3, SFWM.dLambda.HotAtom4, SFWM.dLambda.HotAtom5, SFWM.dLambda.HotAtom6}. In the counter-propagating scheme, the pump field and anti-Stokes photons propagate in one direction and the coupling field and Stokes photons propagate in the opposite direction. Furthermore, the propagation directions of the fields and single photons can have a small angle separation. The phase mismatch, $\Delta\phi$, is given by
\begin{equation}
	\Delta\phi = L \left( \vec{k}_p - \vec{k}_{as} + \vec{k}_c - \vec{k}_s \right) \cdot \hat{z},
\end{equation}
where $L$ is the length of the medium, and $\vec{k}_p$, $\vec{k}_c$, $\vec{k}_{as}$, and $\vec{k}_s$ are the wave vectors of the pump and coupling fields and the anti-Stokes and Stokes photons. Suppose the pointing stabilities or uncertainties and beam divergence of the fields and single photons are all $\pm$0.1$^\circ$. With $L = 75$ mm in this study, it is estimated that the average value of $\Delta\phi$ within $\pm$0.1$^\circ$ in the all-copropagating scheme is 0.91 rad, and that in the counter-propagating scheme is 3.7 rad. Based on the theoretical predictions, the generation rate (or the biphoton temporal width) of $\Delta\phi = 0.91$ rad is about 14 times greater (or 2.5 times longer) than that of $\Delta\phi = 3.7$ rad. Therefore, the all-copropagating scheme employed in our biphoton source is the key of a high generation rate.

Although the all-copropagating scheme was utilized in our previous work of Ref.~\onlinecite{OurOPEX2021} and in this work, the emphases of the previous work and this work are completely different. In Ref.~\onlinecite{OurOPEX2021}, we focused on the biphoton linewidth. Here, we focused on the generation rate of the biphoton source. The all-copropagating scheme is the key ingredient for the study of the generation rate in this work, but it is just a good experimental arrangement for the study of the linewidth in the previous work. Furthermore, the enhancement of the generation rate by the pump power usually decreases SBR. In this work, we systematically studied the effect of the atomic vapor temperature, i.e., the optical depth (OD), on the biphoton source, which was not done in Ref.~\onlinecite{OurOPEX2021}. Our study revealed that the OD played an important role to increase both generation rate and SBR together. 

At a spectral brightness of 3.8$\times$ $10^5$ pairs/s/MHz, our biphoton source had a signal-to-background ratio (SBR) of 2.7, which violated the Cauchy-Schwartz inequality for classical light by about 3.6 folds. The spectral brightness can be further enhanced by the pump power. However, such an enhancement by increasing the pump power will make biphotons no longer be nonclassical. Nevertheless, our systematic study showed that both of the present spectral brightness and SBR can be further enhanced by optimizing the vapor cell temperature and pump power. A hot-atom SFWM biphoton source, with a spectral brightness close to the ultimate limit and a sufficiently large SBR, is feasible.

\section{Experimental Setup}

We carried out the experiment with a glass cell containing only $^{87}$Rb atomic vapor. The inner wall of the cell is coated with paraffin. Two strong fields, pump and coupling, of the $p$ and $s$ polarizations were applied to the cell. Biphotons, i.e., pairs of anti-Stokes and Stokes photons, of the $s$ and $p$ polarizations were generated via the process of spontaneous four-wave mixing (SFWM) as depicted in Fig.~\ref{fig:one}(a). In the SFWM process, the pump field and anti-Stokes photon form one Raman transition and the coupling field and Stokes photon form another. The spontaneous decay rates, denoted as $\Gamma$, of excited states $|3\rangle$ and $|4\rangle$ differ by 5.5\%. For simplicity, we used the average value of $\Gamma = 2\pi\times$5.9 MHz in this study.

The pump (or coupling) field comes from an external-cavity diode laser of Toptica DL DLC pro 780 (or 795). The $e^{-2}$ full width of the pump field is 1.4 mm and that of the coupling field is 1.5 mm. A hyperfine optical pumping (HOP) field was employed in this experiment, although its transition is not shown in Fig.~\ref{fig:one}(a). The frequency of the HOP field is red-detuned from the transition of $|5S_{1/2},F=1\rangle$ $\rightarrow$ $|5P_{3/2},F=2\rangle$ by 80 MHz. We set the power of the HOP field to 9 mW, which is strong enough to empty the population in the state of $|5S_{1/2},F=1\rangle$ driven by the coupling field. The HOP field has the hollow-shaped beam profile, while the biphoton is generated in its hollow region. Hence, the HOP-induced decoherence rate in the SFWM process is negligible. 

The pump and coupling fields and the Stokes and anti-Stokes photons all propagate in the same direction as shown in Fig.~\ref{fig:one}(b). After the vapor cell, we first install a dichroic mirror to separate 780 nm light and 795 nm light. Then, the pump and coupling fields are attenuated by polarization filters and etalons. The polarization filter consists of a quarter-wave plate, a half-wave plate, and a polarizer. We put two etalons in series after the polarization filter. According to the measurements in which the vapor cell was moved away, the leakages of the pump and coupling fields merely contribute 18 and 1.4 counts/s/mW to the anti-Stokes and Stokes single-photon counting modules (SPCMs), respectively. The two SPCMs have the dark-count rates of 140$\pm$5 and 220$\pm$20 counts/s. 

We estimated the collection efficiencies of anti-Stokes and Stokes photons by using two laser beams, signal and probe, respectively. The signal (or probe) beam has the same frequency and polarization as the anti-Stokes (or Stokes) photons. The $e^{-2}$ full widths of these two beams are 0.6 mm. The signal (or probe) beam attenuates to 13\% (or 20\%) after passing through all optics including the polarization filter and two etalons. Considering the quantum efficiency of the anti-Stokes (or Stokes) SPCM, the overall collection efficiency of anti-Stokes (or Stokes) photons is 8.4\% (or 13\%). 

The net linewidth of the two etalons in series for the anti-Stokes (or Stokes) photons was about 46 (or 43) MHz, and the photons' transmission through the etalons was kept to 56$\pm$2\% (or 62$\pm$2\%) throughout this work. The precision of all the etalons' temperature controllers is 1 mK, corresponding to the change of etalon's resonance frequency of about 3 MHz. Due to the drift of the ambient temperature, we typically needed to tune the etalon's resonance frequency once per 4 hours.

The important elements, key issues, and new changes of the experimental setup have been described in this section. Other details can be found in our previous work of Ref.~\onlinecite{OurOPEX2021}.

\section{Results and Discussion}

We systematically studied the generation rate and linewidth of biphotons and the background counts as functions of the pump power, $P_{\rm pump}$, and the vapor cell temperature, $T_{\rm cell}$, [i.e., equivalently the optical depth (OD) of the atomic vapor]. $P_{\rm pump}$ was varied from 2 to 16 mW, and $T_{\rm cell}$ ranged between 38 and 65 $^\circ$C. We kept the coupling power to 4.0 mW throughout the study, which corresponds to the Rabi frequency of 5.4$\Gamma$ as determined by the EIT spectrum \cite{HotREIT, HOTEIT, OurOPEX2021}. This coupling Rabi frequency enables the biphoton linewidths measured here to be always around 1 MHz, which is well below the natural linewidths of the atomic transitions. 

\FigTwo

Representative data of the biphoton wave packet in Figs.~\ref{fig:two}(a) and \ref{fig:two}(b) were taken at the highest and lowest vapor cell temperatures in this study. The biphoton wave packet is the measurement of two-photon correlation function, i.e., coincidence count of anti-Stokes and Stokes photons as a function of the delay time between the two photons. In Fig.~\ref{fig:two}(a), the highest generation rate of 3.7$\times$10$^5$ pairs/s of our biphoton source was achieved at the maximum pump power. To calculate the generation rate, we divided the detection rate by the total collection efficiency (1.1\%) of the photon pair. The linewidth of the biphoton wave packet is 960 kHz, and its determination method will be illustrated in the paragraph consisting of Eq.~(\ref{eq:fitting}). The high generation rate plus narrow linewidth results in a spectral brightness or generation rate per linewidth of 3.8$\times$10$^5$ pairs/s/MHz, which, to our knowledge, is the highest among all biphoton sources, including both of the SFWM and SPDC generation processes.

As compared with Fig.~\ref{fig:two}(a), Fig.~\ref{fig:two}(b) shows that with the same pump power the generation rate is greatly reduced and the linewidth of biphoton wave packet significantly increases at the lowest temperature. Furthermore, the success probability of detecting a Stokes photon (excluding the background) upon an anti-Stokes trigger in (a) is 3.4\% and that in (b) is 2.0\%. A higher vapor cell temperature, i.e., a larger OD, helps the success probability. Nevertheless, the signal-to-background ratio (SBR) of 12 in (b) is much better than that of 3.1 in (a), where the SBR is the ratio of amplitude (equal to subtract baseline from peak) count to baseline (i.e., background) count in the data of biphoton wave packet. A higher biphoton generation rate is usually accompanied with a smaller SBR, which has been systematically studied and will be presented later.

\FigThree

At the highest generation rate achieved in this work, we also measured the auto-correlation function of the anti-Stokes photons, $g^{(2)}_{as,as} (\tau)$, and that of the Stokes photons, $g^{(2)}_{s,s} (\tau)$ as shown in Figs.~\ref{fig:newthree}(a) and \ref{fig:newthree}(b), where $\tau$ is the delay time between two single photons that contributed a coincidence count in the auto-correlation measurement. The solid lines are the best fits, revealing that the anti-Stokes photons had $g^{(2)}_{as,as}(0) =$ 1.95$\pm$0.03 with a full width at the half maximum (FWHM) of 150~ns, and the Stokes photons had $g^{(2)}_{s,s}(0) =$ 1.97$\pm$0.07 with a FWHM of 140~ns. On the other hand, the biphoton wave packet of the highest generation rate exhibited a SBR (denoted as $R_{\rm SB}$) of 2.7$\pm$0.3 with a FWHM of 180 ns, while the representative data in Fig.~\ref{fig:two}(a) had a little higher SBR of 3.1. The cross-correlation function between the anti-Stokes and Stokes photons, $g^{(2)}_{as,s}(0)$, is given by \cite{SBRvsCrossg2}
\begin{equation}
		g^{(2)}_{as,s}(0) = 1 + R_{\rm SB}.
\end{equation}%
Therefore, the measured values of $g^{(2)}_{as,s}(0)$, $g^{(2)}_{as,as}(0)$, and $g^{(2)}_{s,s}(0)$ reveal that the Cauchy-Schwartz inequality for classical light, i.e., $[g^{(2)}_{as,s}(0)]^2/[g^{(2)}_{as,as}(0) \cdot g^{(2)}_{s,s}(0)] \leq 1$, is violated by 3.6 folds. The nonclassicality of these high-rate and narrow-linewidth biphotons has been clearly demonstrated.

Theoretical predictions of the two-photon correlation function, $G^{(2)}(\tau)$, are calculated with the formulas shown below \cite{SFWM.Theory}:
\begin{eqnarray}
\label{eq:biphoton}
	G^{(2)}(\tau) \!\!&=&\!\! \left| 
		\int_{-\infty}^{\infty} d\delta \frac{e^{-i\delta\tau}}{2\pi}
		\bar{\kappa}(\delta) 
		\, {\rm sinc} \!\left[ \bar{\rho}(\delta) \right]
		\exp \!\left[ i \bar{\rho}(\delta) \right]
		\right|^2, 
		\\
\label{eq:kappa}
	\bar{\kappa}(\delta) \!\!&=&\!\!
		\int_{-\infty}^{\infty} d\omega_D  
		\frac{e^{-\omega_D^2/\Gamma_D^2}}{\sqrt{\pi}\Gamma_D}
		\left[ 
		\frac{\alpha\Gamma}{4} 
		\frac{\Omega_p}{\Delta_p-\omega_D + i\Gamma/2}
		\right. \nonumber \\
	&\times&\!\! \left.
		\frac{\Omega_c}{\Omega_c^2-4(\delta+i\gamma)(\delta+\omega_D+i\Gamma/2)}
		\right],	
		\\
\label{eq:rho}
	\bar{\rho}(\delta) \!\!&=&\!\!
		\int_{-\infty}^{\infty} d\omega_D  
		\frac{e^{-\omega_D^2/\Gamma_D^2}}{\sqrt{\pi}\Gamma_D}
		\nonumber \\
	&\times&\!\!
		\left[ 
		\frac{\alpha\Gamma}{2} 
		\frac{\delta+i\gamma}{\Omega_c^2-4(\delta+i\gamma)(\delta+\omega_D+i\Gamma/2)}
		\right],
\end{eqnarray}
where $\tau$ is the delay time between the anti-Stokes and Stokes photons, $\alpha$ represents the OD of the entire atoms, $\Omega_p$ and $\Omega_c$ are the Rabi frequencies of the pump and coupling fields, $\Gamma_{D}$ is the Doppler width, $\Delta_p$ is the pump detuning, and $\gamma$ is the decoherence rate. The other parameters in the above equations are irrelevant to the experimental condition, and their physical meanings can be found in Ref.~\onlinecite{OurOPEX2021}. In all the theoretical calculations, we set $\Gamma_{D}$ to a fixed value of 55$\Gamma$, corresponding to the mediate temperature in the study. A 5\% change in $\Gamma_D$ makes a negligible difference in the calculation result.

The red lines in Figs.~\ref{fig:two}(a) and \ref{fig:two}(b) are the theoretical predictions. In the theoretical calculation, we set the experimentally-determined values of $\alpha$ and $\Omega_c$, and varied $\gamma$ in a range which is reasonable according to the EIT spectrum \cite{HOTEIT, OurOPEX2021}. The values of $\Delta_p$ and $\Omega_p$ do not affect the temporal profile calculated from Eq.~(\ref{eq:biphoton}). The agreement between the experimental data and theoretical predictions is satisfactory.

The green lines in Figs.~\ref{fig:two}(a) and \ref{fig:two}(b) are the best fits. We employ the following phenomenological function in the fitting:
\begin{eqnarray}
	&&
		\!\!\!\!\! \left\{ A \left[ 1 + \tanh\left( \frac{t-t_0}{\tau_1} \right) \right]^p +\epsilon \right\}
	 	\left[ 1 - {\rm erf}\left( \frac{t-t_0 - t_d}{\tau_2} \right) \right]
		\nonumber \\
	&& 
		+B,
\label{eq:fitting}
\end{eqnarray}
where ${\rm erf}(\,)$ is the error function, and $A$, $B$, $\epsilon$, $t_0$, $p$, $\tau_1$, $\tau_2$, and $t_d$ are the fitting parameters. The fittings are rather successful as shown by the two figures. We then perform the numerical Fourier transform of the square root of the best fit, and square the transform result to obtain the frequency spectrum as shown in the inset. The spectral profile of the biphoton wave packet is nearly a Lorentzian function, and its full width at the half maximum (FWHM) gives the linewidth.

We systematically measured the generation rate of biphotons as a function of $P_{\rm pump}$ at different $T_{\rm cell}$'s as shown in Fig.~\ref{fig:three}(a). The experimental data are fitted with a straight line of zero interception. All of the best fits are in good agreement with the data. The good agreement demonstrates that the generation rate is linearly proportional to the pump power. This is expected from the formula of two-photon correlation function, $G^{(2)}(\tau)$, in Eq.~(\ref{eq:biphoton}), since the generation rate is given by $\int G^{(2)}(\tau) d\tau$ and $G^{(2)}(\tau) \propto \Omega_p^2$. In Fig.~\ref{fig:three}(a), one can also see that at a given pump power a higher cell temperatures, i.e., equivalent a larger vapor pressure or optical depth of the atoms, results in a higher generation rate. During the measurement of Fig.~\ref{fig:three}(a), the detection (or generation) rate of our biphoton source fluctuated $\pm$4\% in 4 hours without tuning the etalons. As we set the experiment to the same condition, the day-to-day fluctuation of the detection rate was about $\pm$6\%.

To study how the generate rate quantitatively depends on the OD, we plot the best fit at $P_{\rm pump}$ of 16 mW determined in Fig.~\ref{fig:three}(a) against the measured OD, $\alpha'$. The data points are shown by the circles in Fig.~\ref{fig:three}(b). The value of $\alpha'$ was equal to the minus logarithm of the transmission at the resonance frequency of $|1\rangle$ $\rightarrow$ $|3\rangle$, and it was determined by the fitting of one-photon absorption spectrum of a probe field. Hence, $\alpha'$ relates to $\alpha$ used in Eqs.~(\ref{eq:biphoton})-(\ref{eq:rho}) as the following:
\begin{equation}
\label{eq:OD}
	\alpha' = \alpha \frac{\sqrt{\pi}}{2} \frac{\Gamma}{\Gamma_D}.
\end{equation}
We calculate the generation rate, i.e., $\int G^{(2)}(\tau) d\tau$, as a function of $\alpha$, and then convert $\alpha$ to $\alpha'$ by using Eq.~(\ref{eq:OD}). In the calculation, $\gamma$ is set to the mean value of the decoherence rates determined in Figs.~\ref{fig:two}(a) and \ref{fig:two}(b). The black line in Fig.~\ref{fig:three}(b) is the theoretical prediction of generation rate against $\alpha'$. The consistency between the theoretical prediction and experimental data is acceptable.

\FigFour

In the measurement of Fig.~\ref{fig:three}(a), we also recorded the number of Stokes-photon triggers per second, i.e., the trigger rate $R_t$. The data show $R_t$ is linearly proportional to $P_{\rm pump}$, i.e., $R_t = k_t P_{\rm pump}$. We fit the data, and obtain $k_t \approx$ (1.8, 2.6, 4.1, 6.2, and 7.6)$\times$ $10^3$ counts/s/mW corresponding to the five $T_{\rm cell}$'s from low to high, respectively. Furthermore, $R_t$ is linearly proportional to the OD, i.e., $k_t \approx $ (1.25$\times$ $10^3$)$\alpha'$ counts/s/mW. The linear dependences of the trigger rate on the pump power and optical depth are the expected results of the two-photon Raman transition.

The background count or the baseline in a biphoton wave packet, e.g., 36 counts/bin during the accumulation time of 120~s in Fig.~\ref{fig:two}(b), can be contributed from the four major sources: the SPCM's dark counts, leakage of the coupling field, fluorescence photons induced by the coupling field, and untriggered Stokes photons. The untriggered Stokes photon, whose corresponding anti-Stokes photon of the same pair is unable to make a trigger, contributes a count due to being detected by the SPCM, after a trigger is made by the anti-Stokes photon of a different pair. We use the trigger rate of Fig.~\ref{fig:two}(b) to estimate the background counts contributed from the first three sources. Since the SPCM's dark counts and the coupling leakage produce the backgrounds of 0.60 and 0.015 counts/bin per 120~s, respectively, they are completely negligible. Besides, the HOP field optically pumps the population out of the ground state driven by the coupling field, and it was employed in the experiment. The fluorescence photons, induced by the coupling field and transmitting through the polarization and etalon filters, contributed 1.4 counts/bin per 120~s, which is also negligible. Thus, the background of a biphoton wave packet is mostly contributed by the untriggered Stokes photons. 

To verify that the counts of untriggered Stokes photons dominate the background, we plot the background count rate per time bin as a function of the pump power square ($P_{\rm pump}^2$) in Fig.~\ref{fig:four}(a). Biphotons or pairs of anti-Stokes and Stokes photons are produced randomly in time. The rate of two pairs produced within a given time window, $\Delta t$, is proportional to $[\int G^{(2)}(\tau) d\tau]^2 \Delta t$, where $\Delta t$ for example is 1.6~$\mu$s in Fig.~\ref{fig:two} and $G^{(2)}(\tau)$ is given by Eq.~(\ref{eq:biphoton}). As one of these two pairs makes a trigger and the other contributes a coincidence count, this coincidence count of the untriggered Stokes photon becomes the background Since $G^{(2)}(\tau) \propto P_{\rm pump}$, the background count rate is linearly proportional to $P_{\rm pump}^2$. We fit each data set of a given $T_{\rm cell}$ with a straight line with a nonzero interception. The nonzero interception is resulted from the leakage and fluorescence produced by the coupling field and the SPCM's dark counts. The consistency between the straight line and data demonstrates that the background is predominately contributed from the counts of untriggered Stokes photons.

\FigFive

There is another evidence showing that the rate of two biphoton pairs generated within the triggered time window of 1.6~$\mu$s was significant. Similar to the trigger rate ($R_t$) mentioned previously, the anti-Stokes photon count rate ($R_s$) behaves like $R_s = k_s P_{\rm pump}$, where $k_s \approx$ (1.9, 2.9, 4.6, 7.2, and 9.2)$\times$ $10^3$ counts/s/mW corresponding to the five $T_{\rm cell}$'s from low to high, respectively. Thus, $R_s$ is significantly larger than $R_t$, indicating it happened many times that the anti-Stokes photon in the first pair contributed a count to make a trigger, and that in the second pair contributed another count during the triggered time window. As expected, the difference between $R_s$ and $R_t$ is also linearly proportional to $P_{\rm pump}^2$. 

In Fig.~\ref{fig:four}(b), we show the background count rate per time bin as a function of the measured OD. The black circles are the best fits at $P_{\rm pump}$ of 16 mW determined in Fig.~\ref{fig:four}(a). The black line is the theoretical prediction, which is proportional to $[\int G^{(2)}(\tau) d\tau]^2$ as mentioned earlier. We use the same values of $\Omega_c$, $\gamma$, and $\alpha$ (or equivalently $\alpha'$) to calculate the prediction as those used in Fig.~\ref{fig:three}(b). The agreement between the prediction and data demonstrates again that the background is governed by the counts of untriggered Stokes photons.

The singal-to-background ratio (SBR) of the biphoton wave packet relates to the ratio of the generation rate to the background count rate. Figure~\ref{fig:three}(a) experimentally verified that the biphoton generation rate is proportional to the pump power ($P_{\rm pump}$). Figure~\ref{fig:four}(a) experimentally verified that the background count rate is proportional to $P_{\rm pump}^2$ with some small interception, where the interception was the contribution from the sources other than the untriggered Stokes photons. Thus, one can predict that the SBR as a function of $P_{\rm pump}$ should behave like 
\begin{equation}
\label{eq:fit5a}
	{\rm SBR} = A_1 \frac{P_{\rm pump}}{P_{\rm pump}^2+B_1}, 
\end{equation}
where $A_1$ and $B_1$ are constants independent of $P_{\rm pump}$. In Fig.~\ref{fig:five}(a), the circles represent the experimental data of SBR as a function of $P_{\rm pump}$ taken at the five temperatures, and the lines are the best fits of the above function of $P_{\rm pump}$. The well fittings verify the prediction. While Fig.~\ref{fig:three}(a) shows that either a larger pump power or a higher temperature, i.e. a larger OD, makes a larger generation rate, Fig.~\ref{fig:five}(a) reveals that it causes a smaller SBR.

\FigSix

For each biphoton data as the examples shown in the main plots of Fig.~\ref{fig:two}, we determine the spectral linewidth by the procedure illustrated in the discussion of Fig.~\ref{fig:two}. The circles in Fig.~\ref{fig:five}(b) are the experimental data of spectral linewidth as a function of $P_{\rm pump}$. According to Eq.~(\ref{eq:biphoton}), the frequency spectrum, $F(\delta)$, of the biphoton wave packet is given by
\begin{equation}
\label{eq:spectrum}
	F(\delta) = \left| \bar{\kappa}(\delta) \, {\rm sinc} \!\left[ \bar{\rho}(\delta) \right]
		\exp \!\left[ i \bar{\rho}(\delta) \right] \right|^2,
\end{equation}
where $\delta$ is the two-photon detuning of the Raman transition between the Stokes photon and coupling field, or equivalently the frequency deviation of the Stokes (or anti-Stokes) photon with respect to the resonance. In principle, the linewidth of $F(\delta)$ is independent of the pump Rabi frequency $\Omega_p$ or $P_{\rm pump}$ based on $\bar{\kappa}(\delta)$ and $\bar{\rho}(\delta)$ in Eqs.~(\ref{eq:kappa}) and (\ref{eq:rho}). However, the presence of the pump field can slightly increase the decoherence rate, $\gamma$, in the system \cite{QM2}, which is not considered in Eq.~(\ref{eq:biphoton}). A larger $P_{\rm pump}$ results in a larger $\gamma$, and a larger $\gamma$ in $F(\delta)$ produces a larger linewidth \cite{HotAtomBiphotonLinewidth}. Consequently, the spectral linewidth of the biphotons slightly increases with the pump power, and the data in Fig.~\ref{fig:five}(b) behave like 
\begin{equation}
\label{eq:fit5b}
	{\rm Spectral~Linewidth} = A_2 P_{\rm pump} + B_2, 
\end{equation}
where $A_2$ and $B_2$ are constants independent of $P_{\rm pump}$.

The biphoton linewidth decreases with the OD or the vapor cell temperature as shown by Fig.~\ref{fig:five}(b). This is expected from Eqs.~(\ref{eq:rho}) and (\ref{eq:spectrum}). A larger OD results in a narrower EIT bandwidth of $|\!\exp[i\bar{\rho}(\delta)]|^2$, and plays a negligible role in the bandwidth of $|\bar{\kappa}(\delta) \; {\rm sinc}[ \bar{\rho}(\delta)]|^2$. Thus, one can increase the vapor cell temperature, i.e., the OD, to enhance the generation rate, and also get the bonus of narrowing the linewidth.
 
The generation rate per linewidth, i.e., the spectral brightness, is an important figure of merit of a biphoton source. The circles in Fig.~\ref{fig:five}(c) are the experimental data of the spectral brightness as a function of $P_{\rm pump}$. Considering the generation rate and the linewidth as functions of $P_{\rm pump}$ discussed earlier, the spectral brightness should behave like 
\begin{equation}
\label{eq:fit5c}
	{\rm Spectral~Brightness} = A_3 \frac{P_{\rm pump}}{P_{\rm pump} + B_3}, 
\end{equation}
where $A_3$ and $B_3$ are constants independent of $P_{\rm pump}$. The lines in Fig.~\ref{fig:five}(c) are the best fits with the above function. The well fittings clearly demonstrate the behavior of spectral brightness. As mentioned in the introduction section, the spectral brightness should have an upper limit to make biphotons well separated, and the limit is about $(\pi/2)$$\times$10$^6$ pairs/s/MHz. In Fig.~\ref{fig:five}(c), the highest spectral brightness is 3.8$\times$10$^5$ counts/s/MHz, which is about one quarter of this limit.

Both a larger spectral brightness and a larger signal-to-background ratio (SBR) make a higher success rate in quantum communication. However, a larger spectral brightness always accompanies a smaller SBR as shown by Figs.~\ref{fig:five}(a) and \ref{fig:five}(c), no matter whether the pump power or the vapor cell temperature, i.e., the OD, is varied. To study the spectral brightness and SBR together, we define the quantity $S$ which is the product of spectral brightness and SBR. The circles in  Fig.~\ref{fig:five}(d) represent the experimental data of $S$ as a function of $P_{\rm pump}$. The data show that $S$ initially increases with $P_{\rm pump}$ and approaches an asymptotic limit at $P_{\rm pump} >$ 4 mW. A higher temperature or a larger OD results in a larger asymptotic limit of $S$.

We illustrate the behavior of $S$ versus $P_{\rm pump}$ as follows. The ``signal" of SBR is proportional to the generation rate divided by the biphoton's temporal width, or equivalently the generation rate multiplied by the biphoton's linewidth. On the other hand, the ``background" of SBR is proportional to the background count rate per time bin. Thus, $S$ is proportional to the generation rate square divided by the background count rate per time bin. The generation rate is linearly proportional to $P_{\rm pump}$ as demonstrated in Fig.~\ref{fig:three}(a), while the background count rate per time bin is linearly proportional to $P_{\rm pump}^2$ with a small interception as demonstrated in Fig.~\ref{fig:four}(a). Thus, we can expect that $S$ as a function of $P_{\rm pump}$ behaves like 
\begin{equation}
\label{eq:fit5d}
	S = A_4 \frac{P_{\rm pump}^2}{P_{\rm pump}^2 + B_4}, 
\end{equation}
where $A_4$ and $B_4$ are constants independent of $P_{\rm pump}$. The lines in Fig.~\ref{fig:five}(d) are the best fits with the above function. The well fittings verify the behavior. Furthermore, we plot the SBR against the spectral brightness in Fig.~\ref{fig:seven}. It can be clearly seen that at a given vapor cell temperature a larger spectral brightness results in a smaller SBR, and the product of the two is approximately a constant as the pump power becomes large.

The biphoton wave packet at the spectral brightness of 3.8$\times$10$^5$ pairs/s/MHz had the SBR of 2.7. A larger value of SBR is desirable, e.g., a SBR of 10 indicates that the probability of the single-photon event is 90\%. However, as shown by Fig. 5(d), we operated the biphoton source in the regime that $S$ is nearly constant. Further increasing the SBR will result in decreasing the spectral brightness. Nevertheless, it is also indicated by Fig.~5(d) that a larger optical depth, i.e., a higher atomic vapor temperature, can further enhance the product of spectral brightness and SBR. Suppose the present vapor cell could be heated to above 65~$^\circ$C, which was the highest temperature in this work due to the paraffin coating of the cell. Then, a spectral brightness around the present best value with a SBR better than 10 can be achievable.

An interesting regime below the threshold of an optical parametric oscillator was studied in Ref.~\cite{OPO}. In the regime that the biphoton model is no longer valid, the nonlinear dependence of pump power and the linewidth narrowing of biphotons were observed. To see whether Eq.~(\ref{eq:biphoton}) based on the biphoton model can still be used in the present study, we examined the experimental data and condition. The data in Fig.~\ref{fig:three}(a) reveal that the generation rate of the biphotons was linearly proportional to the pump power, and there was no observable sign of nonlinearity even at the highest pump power and the largest optical depth. As for the largest generation rate achieved in this study, the average separation between two biphotons was about 16 times longer than their temporal width, indicating almost all of the biphotons were away from each other. The data in Fig.~5(b) show that the linewidth of biphotons was slightly broadened as the pump power increased, and there was no hint of linewidth narrowing. Besides, no optical cavity was employed in the experiment. In view of the above examinations, it is certain that Eq.~(\ref{eq:biphoton}) is valid for the experimental data in this study.

\FigSeven

\section{Conclusion}

Utilizing the all-copropagating scheme in the SFWM process, biphotons were generated from a paraffin-coated glass cell containing a hot vapor of nearly all $^{87}$Rb atoms. The highest generation rate of our biphoton source was 3.7$\times$ $10^5$ pairs/s, which is an order of magnitude better than the previous best result of all SFWM sources. At the same time, the biphoton wave packet had a temporal width of 180 ns and a spectral linewidth of 960 kHz. The high generation rate together with the narrow linewidth resulted in a spectral brightness of 3.8$\times$ $10^5$ pairs/s/MHz, which is the best record to date among all biphoton sources, including both of the SFWM and SPDC generation processes. The spectral brightness is an important figure of merit of a biphoton source, and strongly influences the success rate of quantum communication. It has an upper limit of ($\pi$/2)$\times$10$^6$ counts/s/MHz. The achieved spectral brightness in this experiment is about one quarter of the upper limit.

In the experiment, the generation rate, the background count rate, the signal-to-background ratio (SBR), and the linewidth as functions of the pump power and the optical depth (OD) have been systematically studied. The functions or behaviors formed by the experimental data are in good agreement with those predicted by the theory. Furthermore, both a high spectral brightness and a large SBR are desirable, but a higher spectral brightness always accompanies a smaller SBR. The results of this study showed that by increasing the pump power the product of spectral brightness and SBR approached an asymptotic limit, and this asymptotic limit increased with the OD. As the spectral brightness of the biphoton source reached 3.8$\times$ $10^5$ pairs/s/MHz, the SBR decreased to 2.7 due to the asymptotic limit at the highest allowed temperature of the vapor cell. As long as a vapor cell can be heated to a higher temperature, both of the spectral brightness and SBR can be enhanced simultaneously. A biphoton source, with a spectral brightness close to the ultimate limit and a sufficiently large SBR, is feasible.

In conclusion, the result reported in this work is an important milestone, showing that the spectral brightness of the SFWM biphotons can approach to the ultimate limit, and be better than that of the SPDC biphotons. The systematic study as well as the data shown in Figs.~5(a)-(d) provides useful knowledge of quantum technology utilizing heralded single photons.

\section*{ACKNOWLEDGMENTS}
We thank the Referee who pointed out the relation between the signal-to-background ratio in biphoton wave packets, and the cross-correlation function between anti-Stokes and Stokes photons. This work was supported by the Ministry of Science and Technology of Taiwan, under Grants No. 109-2639-M-007-002-ASP and No. 110-2639-M-007-001-ASP.

\end{document}